\documentclass[fleqn,twoside]{article}
\usepackage{espcrc2}
\usepackage{amsmath}
\usepackage{amssymb}

\usepackage{graphicx}
\usepackage[figuresright]{rotating}

\newcommand{\stheta}{\sin^22\theta_{13}}
\newcommand{\nova}{NO$\nu$A}

\title{Combined potential of future long-baseline and reactor
  experiments\thanks{Talk given by T.S.\ at the NOW2004 workshop, Conca
  Specchiulla (Otranto, Italy), 11--17 Sept.\ 2004.}}

\author{P. Huber\address{\vspace*{-2mm}Department of Physics, University of
       Wisconsin, 1150 University Avenue, Madison, WI 53706, USA},
       M. Lindner\address[TUM]{\vspace*{-2mm}Physik--Department, TU M\"unchen,
       James--Franck--Strasse, D--85748 Garching, Germany},
       M. Rolinec\addressmark,
       T. Schwetz\address{\vspace*{-2mm}Scuola Internazionale
       Superiore di Studi Avanzati, Via Beirut 2--4, I--34014 Trieste, Italy}
       and W. Winter\address{School of Natural Sciences, Institute for
       Advanced Study, Einstein Drive, Princeton, NJ 08540, USA} }
       
\begin{document}

\begin{abstract}
We investigate the determination of neutrino oscillation parameters by
experiments within the next ten years. The potential of conventional beam
experiments (MINOS, ICARUS, OPERA), superbeam experiments (T2K, \nova), and
reactor experiments (D-CHOOZ) to improve the precision on the ``atmospheric''
parameters $\Delta m^2_{31}$, $\theta_{23}$, as well as the sensitivity to
$\theta_{13}$ are discussed. Further, we comment on the possibility to
determine the leptonic CP-phase and the neutrino mass hierarchy if
$\theta_{13}$ turns out to be large.
\end{abstract}

\maketitle

%\section{INTRODUCTION}

Triggered by the spectacular results in neutrino physics during the previous
ten years~\cite{results}, several new experimental projects are under way in
this field.  In this note we investigate where we could stand in the
determination of neutrino oscillation parameters in ten years from now, by
considering the experiments which are under construction or under discussion
now, but could deliver physics results within the anticipated time scale.  In
particular we consider the conventional beam experiments
MINOS~\cite{Ables:1995wq}, and the CERN to Gran Sasso (CNGS) experiments
ICARUS~\cite{Aprili:2002wx} and OPERA~\cite{Duchesneau:2002yq}, the superbeam
experiments J-PARC to Super-Kamiokande (T2K)~\cite{Itow:2001ee} and NuMI
off-axis (\nova)~\cite{Ayres:2002nm}, as well as new reactor neutrino
experiments~\cite{reactors} with a near and far detector. The main
characteristics of these experiments are given in Tab.~\ref{tab:experiments}.
For the reactor experiments we use the Double-Chooz proposal
(D-CHOOZ)~\cite{doublechooz} as initial stage setup with roughly $6\times
10^4$ events, and an optimized setup called Reactor-II, with a slightly longer
baseline and $6\times 10^5$ events. Such a configuration could be realised at
several other sites under discussion~\cite{reactors}. 
The results presented in the following are based on Ref.~\cite{prospect},
where more details on the analysis are available. The simulation of the
experiments as well as the statistical analysis is performed with the GLoBES
software package~\cite{globes}.

\begin{table}[t]
\caption{Characterisitics of the considered experiments.}
\label{tab:experiments}
\begin{tabular}{l@{\hspace{-0.8ex}}rrrl}
\hline
 Label &  $L\, [\mathrm{km}]$ &  $\langle E_\nu \rangle$ 
& $t_{\mathrm{run}}$ & 
%$m_\mathrm{Det}$ &  
channel \\
\hline
\multicolumn{5}{l}{ \bf{Conventional beam experiments:}}\\
 MINOS\ & $735$ & 
 $3 \,\mathrm{GeV}$ & $5 \, \mathrm{yr}$ & %$5.4 \,\mathrm{kt}$ &
 $\nu_\mu \!\to\! \nu_{\mu,e}$\\
 ICARUS &  $732$ &  
 $17\,\mathrm{GeV}$  & 
 $5 \, \mathrm{yr}$ & %$2.35 \,\mathrm{kt}$ &
 $\nu_\mu \!\to\! \nu_{e,\mu,\tau}$\\
 OPERA &   $732$ &  
 $17\,\mathrm{GeV}$ & 
 $5 \, \mathrm{yr}$ & %$1.65 \,\mathrm{kt}$ &
 $\nu_\mu \!\to\! \nu_{e,\mu,\tau}$\\
\multicolumn{5}{l}{\bf{Off-axis superbeams:}} \\
 T2K &  $295$ & 
 $0.76 \, \mathrm{GeV}$ & 
 $5 \, \mathrm{yr}$ & %$22.5 \,\mathrm{kt}$ &
 $\nu_\mu \!\to\! \nu_{e,\mu}$\\
 \nova\ & 
 $812$ & 
 $2.22 \, \mathrm{GeV}$ & 
 $5 \, \mathrm{yr}$ & %$50 \,\mathrm{kt}$ &
 $\nu_\mu \!\to\! \nu_{e,\mu}$\\
\multicolumn{5}{l}{
\bf{Reactor experiments:}} \\
 D-CHOOZ & 
 $1.05$ & 
 $\sim 4 \, \mathrm{MeV}$ & 
 $3 \, \mathrm{yr}$ & %$11.3 \,\mathrm{t}$ &
 $\nu_e \!\to\! \nu_e$\\
 Reactor-II & 
 $1.70$ & 
 $\sim 4 \, \mathrm{MeV}$ & 
 $5\,\mathrm{yr}$ & %$200 \,\mathrm{t}$ &
 $\nu_e \!\to\! \nu_e$\\
\hline
\end{tabular}
\end{table}

%%%%%%%%%%%%%%%%%%%%%%%%%%%%%%%%%%%%%%%%%%%%%%%%%%%%%%%%%%%%%%%%

%\section{ATMOSPHERIC PARAMETERS} 

\begin{figure*}[t]
\centering
\includegraphics[width=0.75\textwidth]{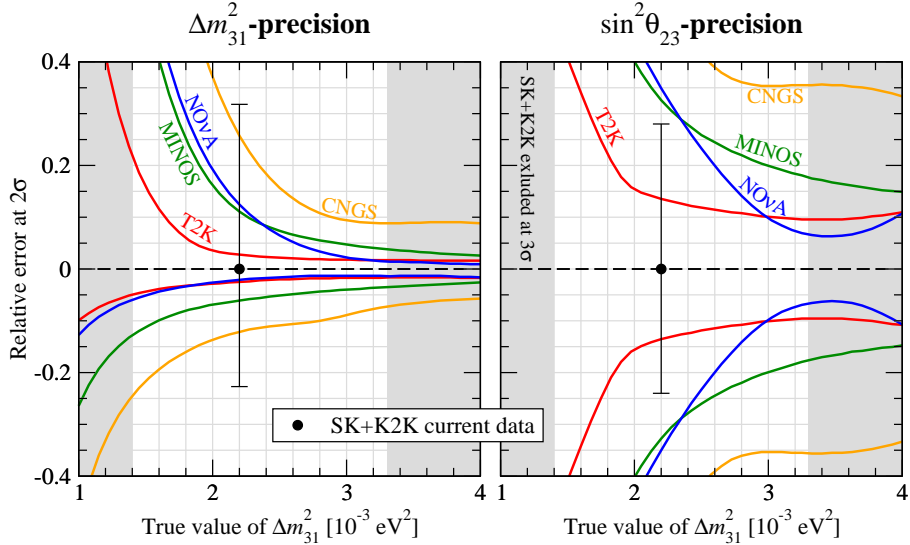}
\vspace*{-1cm}
\caption{Precision of $\Delta m^2_{31}$
  (left) and $\sin^2\theta_{23}$ (right) as a function of the
  true value of $\Delta m^2_{31}$; $\theta_{23}^\mathrm{true} = \pi/4$.}
\label{fig:atm}
\end{figure*}

First we discuss the improvement which can be expected for the ``atmospheric''
parameters $\Delta m^2_{31}$ and $\sin^2\theta_{23}$. In Tab.~\ref{tab:atm} we
show the precision at $3\sigma$, which we define as $(x^{+3\sigma} -
x^{-3\sigma}) / x^0$, where $x^{+(-)3\sigma}$ is the $3\sigma$ upper (lower)
bound, and $x^0$ is the best fit value of the quantity $x$. We compare in the
table the precision of the conventional beams (MINOS+CNGS) and the superbeams
to the current precision, as obtained from a global fit to SK atmospheric and
K2K long-baseline data~\cite{fit}. In the last row we show the precision which
can be obtained by combining all experiments. We observe from these numbers,
that the accuracy on $\Delta m^2_{31}$ can be improved by one order of
magnitude, whereas the accuracy on $\sin^2\theta_{23}$ will be improved only
by a factor two.

\begin{table}[t]
\caption{Precision for $|\Delta m^2_{31}|$ and $\sin^2\theta_{23}$ at
  $3\sigma$ for the true values $\Delta m^2_{31} = 2\times 10^{-3}$~eV$^2$,
  $\sin^2\theta_{23} = 0.5$.}
\label{tab:atm}
\centering
\begin{tabular}{lrr}
\hline
 & $|\Delta m^2_{13}|$  & $\sin^2\theta_{23}$ \\
\hline
current     & 88\% & 79\% \\
\hline
MINOS+CNGS  & 26\% & 78\% \\
T2K         & 12\% & 46\% \\
\nova\        & 25\% & 86\% \\
\hline
Combination &  9\% & 42\% \\
\hline
\end{tabular}
\end{table}

The numbers of Tab.~\ref{tab:atm} depend to some extent on the true value of
$\Delta m^2_{31}$. Therefore, we show in Fig.~\ref{fig:atm} the precision as a
function of this parameter. We observe that for all experiments the
sensitivity suffers for low values of $\Delta m^2_{31}$. For 
$\Delta m^2_{31} \gtrsim 2\times 10^{-3}$~eV$^2$ T2K will provide a precise
determination of $\Delta m^2_{31}$ at the few \% level. Although \nova\ can put
a comparable bound on $\Delta m^2_{31}$ from below, the upper bound is
significantly weaker, and similar to the bound from MINOS. The reason for this
is a strong correlation between $\Delta m^2_{31}$ and $\theta_{23}$, which
disappears only for $\Delta m^2_{31} \gtrsim 3\times 10^{-3}$~eV$^2$.
From the right panel of Fig.~\ref{fig:atm} one can see that for $\Delta
m^2_{31} \sim 2\times 10^{-3}$~eV$^2$ only T2K is able to improve the current
bound on $\sin^2\theta_{23}$. One reason for the rather poor performance on
$\sin^2\theta_{23}$ is the fact that these experiments are sensitive mainly to
$\sin^22\theta_{23}$. This implies that for $\theta_{23} \approx \pi/4$ it is
very hard to achieve a good accuracy on $\sin^2\theta_{23}$, although
$\sin^22\theta_{23}$ can be measured with relatively high
precision~\cite{minakata}.

%\section{THE BOUND ON $\stheta$}

\begin{figure}[tbh]
\centering
\includegraphics[width=0.4\textwidth]{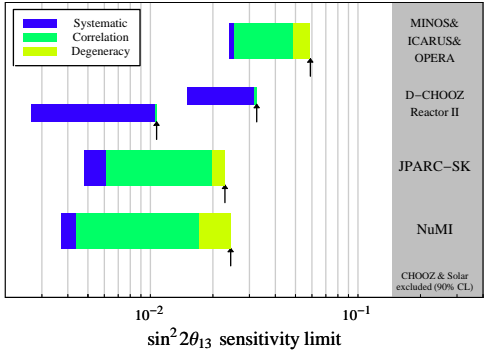}
\vspace*{-0.8cm}
\caption{Sensitivity to $\stheta$ at 90\% CL for the true values 
$\Delta m^2_{31} = 2\times 10^{-3}$~eV$^2$, 
$\Delta m^2_{21} = 7\times 10^{-5}$~eV$^2$.}
\label{fig:th13bars}
\end{figure}

Let us now discuss the sensitivity to $\stheta$, i.e.\ we assume a true value
$\theta_{13} = 0$ and investigate the obtainable upper bound on $\stheta$. The
sensitivities of the various experiments are summarized in
Fig.~\ref{fig:th13bars}, where the impact of systematics, correlations and
degeneracies is indicated by the shading. One immediately observes that the
$\stheta$-limit from beam experiments is strongly affected by parameter
correlations and degeneracies~\cite{rigolin}, whereas reactor experiments
provide a ``clean'' measurement of $\stheta$, dominated by statistics and
systematics~\cite{reactor}. 

\begin{figure}[tbh]
\centering
\includegraphics[width=0.4\textwidth]{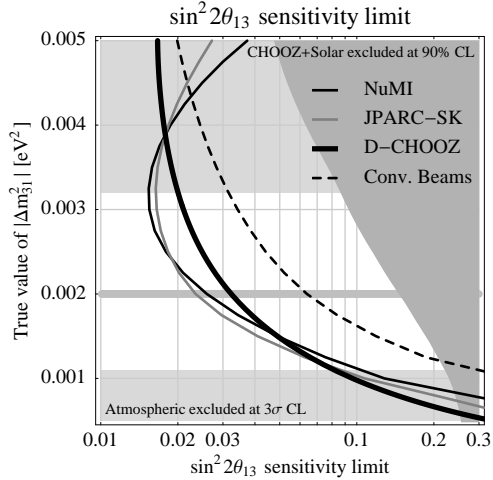}
\vspace*{-1cm}
\caption{Sensitvity to $\stheta$ at 90\%~CL as a function of the true value of
  $\Delta m^2_{31}$.} 
\label{fig:th13}
\end{figure}

The dependence of the $\stheta$-limit on the true value of $\Delta m^2_{31}$
is illustrated in Fig.~\ref{fig:th13}. Again we observe that the sensitivity
of all experiments gets rather poor for low values of $\Delta m^2_{31}$.  For
$\Delta m^2_{31} \sim 2\times 10^{-3}$~eV$^2$ we find roughly an improvement
of a factor 2 from conventional beam experiments (MINOS+ICARUS+OPERA
combined), a factor 4 from D-CHOOZ, and a factor 6 from the superbeams T2K and
\nova\ with respect to the current bound from global data~\cite{fit}. Note
that an optimised reactor experiment such as Reactor-II has the potential for
even better $\stheta$-sensitivities than the superbeams (compare
Fig.~\ref{fig:th13bars}).

%%%%%%%%%%%%%%%%%%%%%%%%%%%%%%%%%%%%%%%%%%%%%%%%%%%%%%%%%%%%%%%%

%\section{POTENTIAL FOR LARGE $\stheta$}

Now we assume a relatively large value $\stheta = 0.1$, close to the current
bound, and investigate what we can learn within the next ten years about the
CP-phase $\delta$ and the neutrino mass ordering.  First we note that all the
experiments will be able to establish the non-zero value.  However, we will be
confronted with allowed regions in the $\theta_{13}$-$\delta$-plane (see
Figs.~8 and 9 of Ref.~\cite{prospect}). None of the experiments on their own
can give any information on the CP-phase $\delta$ and on the mass
hierarchy. The determination of $\stheta$ from beam experiments is strongly
affected by the correlations with $\delta$, and especially for \nova\ also
correlations with other parameters are important. Moreover, the inability to
rule out the wrong mass hierarchy leads to a further ambiguity in the
determination of $\stheta$. In contrast, since the $\bar\nu_e$-survival
probability does not depend on $\delta$, Reactor-II provides a clean
determination of $\stheta$ at the level of 20\% at 90\%~CL. 

If all experiments are combined the complementarity of reactor and beam
experiments allows to exclude up to 40\% of all possible values of the
CP-phase for a given hierarchy. The wrong hierarchy can be ruled out at modest
CL with $\Delta\chi^2 \simeq 3$ due to matter effects in \nova. However, at
high CL still all values of $\delta$ are allowed, and moreover, even for a
given hierarchy CP-conserving and CP-violating values of $\delta$ cannot be
distinguished at 90\%~CL. We add that these results depend to some extent on
the true value of $\delta$.

\begin{figure*}[tbh]
%\vspace{9pt}
\centering
\includegraphics[width=0.75\textwidth]{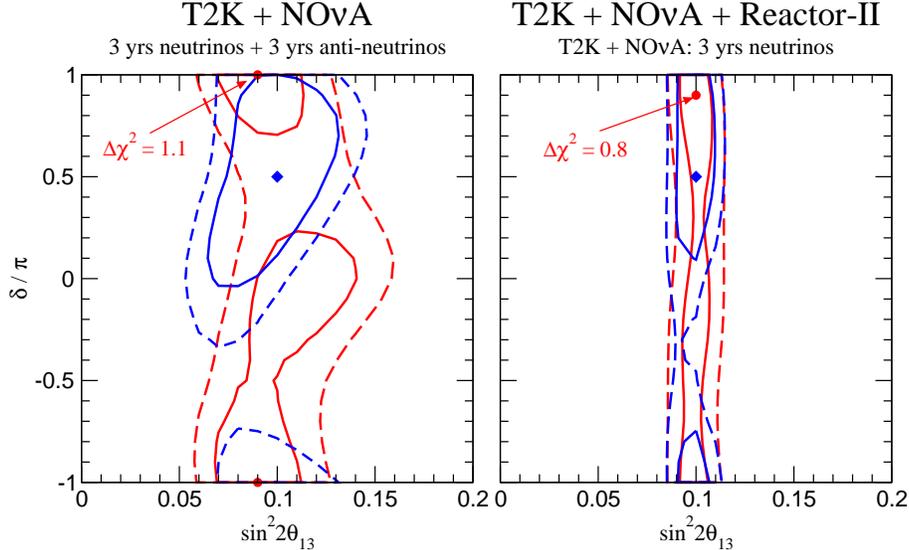}
\vspace*{-1cm}
\caption{Antineutrino running vs Reactor-II. We show the 90\%~CL (solid
   curves) and $3\sigma$ (dashed curves) allowed regions in the
   $\stheta$-$\delta$-plane for the true values $\stheta=0.1$ and
   $\delta=\pi/2$. The blue/dark curves refer to the allowed regions for
   the normal mass hierarchy, whereas the red/light curves refer to the
   $\mathrm{sgn}(\Delta m^2_{31})$-degenerate solution (inverted hierarchy),
   where the projections of the minima onto the $\stheta$-$\delta$-plane are
   shown as diamonds (normal hierarchy) and dots (inverted hierarchy). For the
   latter, the $\Delta\chi^2$-value with respect to the best-fit point is also
   given.}
\label{fig:anti-vs-react}
\end{figure*}

So far we have considered only neutrino running for the superbeams, since it
is unlikely that significant data can be collected with antineutrinos within
ten years from now. Nevertheless, it might be interesting to investigate the
potential of a neutrino-antineutrino comparison. In
Fig.~\ref{fig:anti-vs-react} we show the results from T2K+\nova\ with 3 yrs of
neutrinos + 3 yrs of antineutrinos each (left), in comparison with the case
where the antineutrino running is replaced by Reactor-II (right). We find that
antineutrino data at that level does neither solve the problems related to the
CP-phase nor to the hierarchy. Still CP-violating and CP-conserving values
cannot be distinguished at 90\% CL. Moreover, the determination of $\stheta$
is less precise than from the reactor measurement. To benefit from
antineutrino measurements a significantly longer measurment period would be
necessarry, to obtain large enough data samples.

{\bf Acknowledgement} T.S.\ would like to thank the organizers for a very
pleasant and stimulating workshop. This work has been supported by an
Intra-European Marie Curie fellowship of the EC (T.S.), and by the
``Sonderforschungsbereich 375 f{\"u}r Astro-Teilchenphysik der Deutschen
For\-schungs\-ge\-mein\-schaft''.

\end{document}